\documentstyle[aps,prl,epsf,amsfonts,amssymb]{revtex}
\newcommand{\bu}{{\bf u}}
\newcommand{\bv}{{\bf v}}
\begin{document}
\draft
\twocolumn[\hsize\textwidth\columnwidth\hsize\csname@twocolumnfalse%
\endcsname
\title{One- and two-particle microrheology}
\author{Alex J.~Levine and T. C.~Lubensky}
\address{Department of Physics and Astronomy, University of Pennsylvania,
Philadelphia, PA 19104}

\date{\today}
\maketitle
\begin{abstract}
We study the dynamics of rigid spheres embedded in viscoelastic media and 
address two questions of importance to microrheology.  First we calculate
the complete response to an external force of a single bead in a homogeneous
elastic network viscously coupled to an incompressible fluid.  From this 
response function we find the frequency range where 
the standard assumptions of microrheology are valid.
Second we study fluctuations when embedded spheres perturb the media around
them and show that mutual fluctuations of two separated spheres provide 
a more accurate determination of the complex shear modulus than 
do the fluctuations of a single sphere.
\end{abstract}
]

Microrheology is an important experimental probe of the viscoelastic properties
of soft materials\cite{Mason:95}.  Unlike more traditional macrorheology, 
in which a sample is subjected to an externally imposed 
uniform shear strain, microrheology 
relies on the Brownian fluctuations of the 
micron-sized beads dispersed in the sample to
assess the viscoelastic response function (complex shear modulus), $G(\omega)$. 
The principal advantages of this technique are that it can be used for 
the detailed study of materials that cannot be produced in bulk 
quantities and that it can be used 
to probe the local properties of rheologically inhomogeneous materials.
Because of these two strengths, microrheology promises to open a new window on 
cellular biology by facilitating the study of the rheological properties of 
intra-cellular structures in living cells.  In addition, this technique is 
currently being used to study various soft biomaterials\cite{Schnurr:97}. 

In a typical microrheology experiment, the time-dependent position correlation
function of individual probe particles is measured either by light scattering%
\cite{Schnurr:97} or by direct real--space imaging\cite{Crocker:00}.  This 
correlation function provides a complete description (via the Fluctuation--%
Dissipation Theorem) of the frequency-dependent response of the probe particles
to an external force.  If inertial effects are ignored, the rheological 
properties of a Newtonian fluid are completely determined by a single quantity,
its viscosity $\eta$.  The displacement of a spherical particle, 
$\bu(\omega)$, of radius $a$ in response to a force ${\bf f}(\omega )$ 
at frequency $\omega $ in such a fluid is given by the standard 
Stokes-Einstein relation:
\begin{equation}
\label{Stokes}	
u(\omega) = \frac{f(\omega)}{6 \pi a G(\omega ) }, 
\end{equation}
where $G(\omega ) = -i \omega \eta$ is the complex shear modulus.
A natural hypothesis\cite{Mason:95} is that this relation can be generalized
to rheologically complex materials in which $G(\omega )$, the complex 
shear modulus has  both storage (real) and loss (imaginary) components.
We will refer to this extension of Eq.~(\ref{Stokes}) as the Generalized 
Stokes--Einstein relation (GSER).   

In this Letter we address two basic questions regarding the interpretation of 
microrheological data: (1)   In a homogeneous viscoelastic
medium does the GSER provide the correct response function to an applied force?
and, (2) if the 
introduction of the probe particles perturbs the local rheological properties 
of the medium, how does one extract the unperturbed, bulk rheological 
properties from the data?  Recent experiments suggest
that, at least in certain systems, a discrepancy exits between the 
macro- and microrheological measurement of the shear modulus~%
\cite{Crocker:00} making this question one of current 
interest.  

To address the first of these questions, we use a model viscoelastic 
medium consisting of an elastic network that is viscously coupled to a 
fluid in which the network is embedded \cite{Brochard:77,Schnurr:97}.  
In the second half of this 
letter, we approach the problem of rheological inhomogeneities and explicitly 
show that {\it inter-particle\/} position correlations  are insensitive
to the local particle environment, and therefore, provide
a more reliable probe of the properties of the bulk material than do single
particle fluctuations as proposed in reference \cite{Crocker:00}.

Our model viscoelastic medium consists of an elastic network, characterized
by a displacement variable ${\bf u}$, that is viscously coupled via a friction
coefficient $\Gamma$ to an incompressible Newtonian fluid characterized by a
velocity field ${\bf v}$.  In the absence of viscous coupling, $\bu$ obeys the 
standard equation for an isotropic, elastic, compressible medium with 
Lam\'{e} coefficients $\lambda$, $\mu$,  and $\bv$ obeys the incompressible 
Navier-Stokes equation with viscosity $\eta$.  With friction included,
the equations for $\bu $ and $\bv$%
\cite{Brochard:77} are
\begin{eqnarray} 
\label{ueom}
\rho \ddot{{\bf u}} - \mu \nabla^2 \bu - \left( \lambda +
\mu \right) {\bf \nabla} \left( \nabla \cdot \bu \right) &=& - \Gamma \left( 
\dot{{\bf u}} - \bv \right) + {\bf f}_{\rm u}\\
\label{veom}
\rho_{\rm F} \dot{\bv} - \eta \nabla^2 \bv  + {\bf \nabla} P &=& 
 \Gamma  \left( \dot{\bu} - \bv 
\right) + {\bf f}_{\rm v} \\
\label{incompress}
{\bf \nabla} \cdot \bv &=& 0, 
\end{eqnarray} 
where $P$ is the pressure and ${\bf f}_{\rm u}$ and $ {\bf f}_{\rm v}$ are, 
respectively, the force densities exerted on $\bu$ and $\bv$ by the embedded 
beads.
The friction coefficient $\Gamma $ is estimated by considering a uniform
displacement of the network relative to the fluid at constant relative velocity
$\bv$.  The friction force per unit volume, $\Gamma \bv$, is equal to the friction
force $\eta \xi  \bv$ on a strand of the network of length equal to the mesh size
$\xi  $ divided by $\xi^3$, the volume per strand.  The result is $\Gamma \sim %
\eta /\xi^2$.  

Our goal is to calculate the frequency-dependent displacement compliance 
$\alpha(\omega )$ relating bead displacement ${\bf r}(\omega )$ to the external force
${\bf F}(\omega )$ imposed on it:
\begin{equation}
\label{compliance-definition}
{\bf r}(\omega ) = \alpha(\omega ){\bf F}(\omega ),
\end{equation}
and to determine under which conditions, if any, the GSER, 
$\alpha(\omega )=%
1/\left( 6 \pi a G(\omega ) \right)$ applies, {\it i.e.\/} 
under what conditions 
measurements of the displacement of an 
individual particle provide a direct measure of the complex 
shear modulus of the two-fluid medium.  The complete solution to this 
problem requires solving 
Eqs.~\ref{ueom}--\ref{incompress} with time derivatives replaced by 
$-i \omega $, 
${\bf f}_{\rm u}$ and ${\bf f}_{\rm v}$ equal to zero, and with boundary
conditions that $\bu (\omega ) = \bv/(-i \omega )= {\bf r}(\omega )$ at the surface
of the sphere.  The resulting functions $\bu({\bf x}, \omega )$ and 
$\bv({\bf x}, \omega )$ can then be used to calculate the stress at the surface of
the bead and by integration the total force ${\bf F}_{\rm b}(\omega )$ exerted
on the medium by the bead. Newton's equation for a bead of mass $M$, $- \omega^2
M \bu + {\bf F}_{\rm b}(\omega ) = {\bf F}(\omega )$ then determines $\alpha(%
\omega )$.
This procedure is laborious at best, and we will apply a slightly less 
rigorous one.  We localize the bead--medium forces ${\bf f}_\alpha $, ($\alpha %
= {\rm u, v}$) on the bead by setting ${\bf f}_{\alpha }({\bf k}, \omega ) %
= {\bf F}_{\alpha}(\omega )  \Theta(\left| {\bf k} \right| - k_{\rm max} )$ 
where ${\bf f}_{\alpha } ({\bf k},
\omega )$ is the Fourier transform of ${\bf f}_{\alpha }({\bf x}, t)$, 
${\bf F}_\alpha (\omega )$ is the integrated force exerted by the bead, 
$k_{\rm max} = \pi /2 a $, and $\Theta(x)$ is the unit step function.  The 
total force exerted on the bead by the medium is ${\bf F}_{\rm b}(\omega ) =
{\bf F}_{\rm u}(\omega ) + {\bf F}_{\rm v}(\omega )$ and Newton's equation for
a bead is the same as above.  

Our procedure is to use 
Eqs.~\ref{ueom}--\ref{incompress} to calculate $\bu ({\bf k}, \omega )$ and 
$\bv ({\bf k}, \omega )$ in terms of ${\bf f}_{\rm u} ({\bf k}, \omega )$
and ${\bf f}_{\rm v} ({\bf k}, \omega )$, and then to calculate by integration
over $k$, the network 
displacement ${\bf r}(\omega )$ and fluid velocity ${\bf w}(\omega )$ at the 
bead in terms of ${\bf F}_\alpha(\omega )$.
We then require that the bead, the network, and the fluid all move together at 
the bead, {\it i.e.\/} that ${\bf r}(\omega )$  be the bead displacement and 
${\bf w}(\omega ) = -i \omega {\bf r}(\omega )$ its velocity.  This constraint
on ${\bf r}(\omega ), {\bf w}(\omega )$ imposes a particular ratio between 
${\bf F}_{\rm u}(\omega )$ and ${\bf F}_{\rm v}(\omega )$ that allows us to 
obtain a linear relation between ${\bf r}(\omega )$ and 
${\bf F}_{\rm b}(\omega )$. When applied to a sphere in a Newtonian fluid and 
expanded in powers of $-i \omega $, this procedure reproduces correctly the 
constant and $\sqrt{-i \omega }$ contribution to $\alpha^{-1}(\omega )/(-i 
\omega )$ and the $-i \omega $ inertial contribution with a slightly different
prefactor.  We expect similar accuracy for the current problem.
Our result for $\alpha (\omega )$ can be expressed as 
\begin{equation}
\label{alpha-ans}
\alpha^{-1}(\omega) = \frac{ 6 \pi a G(\omega ) \left( 1- X(\omega) \right) }{
\left[ 1 + H\left( \frac{ \omega}{\omega_{\rm B}} \right) \frac{ G(\omega)}{
2 B} + J(\omega	) \right]} - \omega^2 M
\end{equation}
where we have introduced the complex shear modulus of the material:
$ G(\omega ) = \mu - i \omega \eta$ and the cross-over function, $H$ defined by
\begin{equation}
\label{h-def}
H(x) = 1 - \int_0^1 dz \, \left( 1 + i z^2/x\right)^{-1} 
\end{equation}
as well as the frequency scale: 
$ \omega_{\rm B} = \left( 2 \mu + \lambda \right) / (a^2 \Gamma )$.  
In this result we
assume that the mass density of the elastic network is significantly lower than
that of the fluid, $\rho/\rho_{\rm F} \ll 1$, owing to the open structure of
the network.  Consequently, in Eq.~(\ref{alpha-ans}) we have set $\rho = 0$. 
We have also introduced the functions $J$ and $X$ which we
discuss briefly below.  
A more complete analysis of this result will be 
published elsewhere\cite{Levine:00}.  In order for  the result given by
Eq.~(\ref{alpha-ans}) to reduce to the GSER we must find that (at least for
some frequency range) $H \approx 0$, $ J \approx - X$, and $\beta_{\rm b}
(\omega )= 
\left( 2 \rho_{\rm b} a^2 \omega^2 \right) / 9 G(\omega ) \ll 1$ where 
$ \rho_{\rm b} $ is the mass density of the bead.

First we consider $H$.  From 
Eq.~(\ref{h-def}) we note that $H(x)$ goes to zero as $1/x$ for $x \gg 1$, 
so for
frequencies large compared to $\omega_{\rm B}$, we can neglect this term.  
From an examination of the hydrodynamic modes of the system, the physical 
interpretation of this result is clear.  The frequency scale $\omega_{\rm B}$ 
is the decay time for the over-damped longitudinal compression  mode of 
the system at 
the length scale of the bead.  In this mode the network undergoes a 
compressional wave while the fluid drains from the denser parts of the 
network to the more rarefied parts.  The $H$ function, therefore, represents a
correction to the microrheological measurements due to the excitation of 
longitudinal degrees of freedom in the system.  Whereas in the macrorheological
experiment the applied strain is pure shear, in the microrheological experiment
the probe particle responds to all the thermally excited modes of the system
including the longitudinal compression modes of the elastic network.  
At frequencies 
higher than $\omega_{\rm B}$, however, the network ``locks in'' with 
incompressible fluid thereby eliminating the former's longitudinal modes and
bringing the microrheological measurement into closer correspondence with 
standard rheology. The elimination of the so-called free-draining (longitudinal)
mode at large $\omega $  has been discussed previously
\cite{Schnurr:97}. 

We now consider the function $J(\omega )$.  Its form 
is controlled by two dimensionless parameters: $\beta_{\rm F}(\omega )
 = 4 \omega^2 \rho_{\rm F} a^2/ \left[ G(\omega ) \pi^2 \right] 
$ and $\delta = \left( \xi  / a \right)^2$.  The 
parameter $\beta_{\rm F} $ is formed by the square of the ratio of the sphere's
radius to the inertial decay length\cite{Ferry:80} 
in the medium and measures the importance of fluid inertial effects in the 
compliance.  The second parameter, $\delta $, simply measures the ratio of the 
network mesh size to the sphere radius.  In the limit that both $\beta_{\rm F}$,
 $\delta  \ll 1$ the function $J(\omega	)$ reduces to $-X(\omega) $.  Since
$\rho_{\rm F} \sim \rho_{\rm b}$, $\beta_{\rm b}$ and $\beta_{\rm F}$ are 
of the same order and both will be small for $\omega < \omega^\star$ with 
$\beta_{\rm b}(\omega^\star) \sim \beta_{\rm F}(\omega^\star) \sim 1$.
Our approximate calculation is expected to reproduce the exact result for 
$\omega < \omega^\star$, so our estimate of the region of validity of the 
GSER should be correct.

In typical experiments\cite{Schnurr:97}, the probe 
sphere is taken to be orders of 
magnitude larger than the mesh size so we may safely assume that $\delta \ll 1$.
For experiments on actin \cite{Schnurr:97} with a sphere
size of $1 \, \mu$m,  $\beta_{\rm F} $ remains small up to frequencies on 
the order of $50$kHz.
A similar estimation of the lock-in frequency yields
$\omega_{\rm B} \sim 10$Hz.  
 Thus in typical experiments there remains a significant frequency 
window, $ \omega_{\rm B} < \omega < \omega^\star$, 
where the response function of the probe particle to an applied force
is well approximated by the GSER.  This model calculation reveals the range of
validity of the GSER for typical experiments on soft materials; 
furthermore it presents a 
quantitive prediction of the form of the compliance in frequency regimes where
the GSER does not hold\cite{Levine:00}.

We now turn to the issue of rheological heterogenities introduced by the beads 
themselves.  We imagine a medium characterized by a homogeneous 
frequency-dependent elastic-constant tensor.  The introduction of spherical 
probe particles perturbs the
medium in the vicinity of these particles and leads to a spatially inhomogeneous
elastic constant tensor $K_{ijkl}({\bf x}, \omega )$.  Assuming that the 
stress-strain relation remains local, that the frequency regime 
($\omega_{\rm B} < \omega < \omega^\star$ for our coupled network) is such that 
the medium can be characterized by a single, frequency--dependent elastic 
constant tensor, and that 
inertial terms can be neglected, the equation for the displacement variables is
\begin{equation}
\label{basic-elastic}
- \partial_j \left( K_{ijkl}({\bf x},\omega ) \partial_k u_l \right) = 
f_i({\bf x}, \omega), 
\end{equation}
where $f_i({\bf x},\omega )$ is the force density that acts on the 
surface of the
particles.  The displacement responses of the collection of particles to forces 
upon them can described by a compliance tensor $\alpha_{ij}^{({n m})}$:
\begin{equation}
\label{tensor-alpha}
R_i^{n}(\omega ) = \alpha_{ij}^{({n m})}(\omega ) F_j^{m}(\omega ),
\end{equation}
where $R_i^{n}$ is the displacement vector of the 
$n^{\rm th}$ particle and $F_j^{\rm m}$ the 
force on the $m^{\rm th}$ particle.  We ask which components of the 
compliance tensor depend on the bead-imposed inhomogeneities of 
$K_{ijkl}({\bf x}, \omega )$ and which, if any, depend only on the 
bulk homogeneous part?

To answer this question, it is useful to consider first the simpler but related 
problem of determining the bulk dielectric constant of a medium by measuring the 
self and mutual capacitances of metal spheres whose presence perturbs the 
dielectric constant in their vicinity.  If the dielectric constant $\epsilon({\bf x},%
\omega )$ remains local and frequencies are such that transverse electric fields 
can be ignored, then the potential $\phi({\bf x}, \omega ) $ satisfies
\begin{equation}
\label{electric-potential}
- \nabla \cdot \left( \epsilon({\bf x}, \omega ) \nabla \phi({\bf x}, \omega ) \right) 
= 4 \pi \rho({\bf x}), 
\end{equation}
where $\rho({\bf x})$ is the charge density at ${\bf x}$.  It is clear from 
Eqs.~(\ref{basic-elastic}) and (\ref{electric-potential}) that there is an 
analogy between the electrical and rheological problems with the identification:
$\phi \longleftrightarrow {\bf u}$, $ \epsilon 
\longleftrightarrow K_{ijkl}$, and
$\rho \longleftrightarrow {\bf f}$.  The total charge $Q$ on a metal sphere 
is the analog of the total force ${\bf F}$ on a bead in the viscoelastic medium. 
The inverse capacitance tensor $C^{-1}_{n m }$ defined by
\begin{equation}
\label{capacitance}
\phi_{\rm n} = C^{-1}_{n m } Q_{ m}, 
\end{equation}
where $\phi_{n}$ is the potential on bead $n$ and $Q_{m}$ is the total charge
on bead $m$, is the analog of the compliance tensor.

\begin{figure}
\epsfxsize=0.95\columnwidth
\centerline{\epsfbox{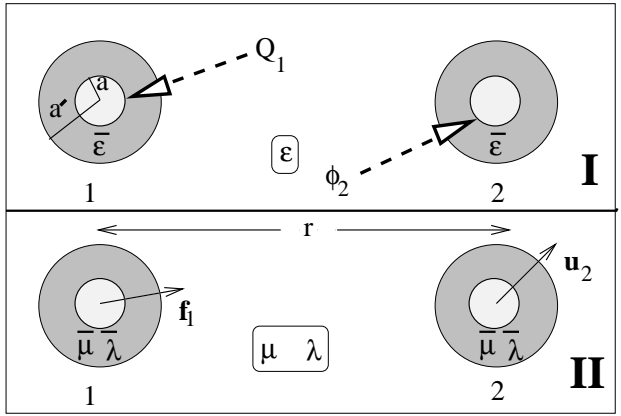}}
\vspace{0.1cm}
\caption[]{
I. Schematic of a system in which conducting spheres of radius $a$ embedded 
in a medium of dielectric constant $\epsilon $ are surrounded by concentric
spheres of radius $a'$ with dielectric constant $\bar{\epsilon }$.  II.
Similar schematic of rigid spheres embedded in an elastic medium with Lam\'{e}
coefficients $\mu$, $\lambda$ surrounded by concentric spheres with 
Lam\'{e} coefficients $\bar{\mu}$, $\bar{\lambda}$.
}
\label{picture}
\end{figure}

To keep our calculation simple, we consider two conducting spheres of radius $a$
separated by a distance $r$ in a medium of dielectric constant 
$\epsilon $.  Each 
sphere perturbs the medium locally, producing a spherical region of 
radius $a'$ with a dielectric constant $\bar{\epsilon }$ as shown in
 Fig.~(\ref{picture}).  To leading order in $a/r$, the inverse 
self-capacitance is
\begin{equation}
\label{self-cap}
C_{11}^{-1} = \frac{ 1}{4 \pi \epsilon a} \left\{ 1 + \left( \frac{ a'}{a} - 1 \right) 
\left( 1 - \frac{ \epsilon }{\bar{\epsilon }} \right) \right\}.
\end{equation}
This result shows that fluctuations of a single
bead are sensitive to the local environment around the bead and 
that, therefore, they do not measure directly the bulk dielectric constant, 
$\epsilon$.  The inverse mutual capacitance, 
\begin{equation}
\label{mutual-cap}
C_{12}^{-1} = \frac{ 1}{4 \pi \epsilon r} \left( 1 + {\cal O} \left( \frac{ a}{r} 
\right)  \right),  
\end{equation}
depends, however, only on the bulk dielectric constant to leading order in $a/r$.  Thus
correlated voltage fluctuations $\langle \phi_1(\omega )\phi_2(- \omega ) \rangle = 
2 \left( T/\omega \right) \, {\rm Im} C^{-1}_{12}(\omega )$ yield 
a direct measurement 
of $\epsilon (\omega )$
provided the beads are far enough apart that $C_{12}^{-1}$ is proportional to $1/r$.

Given the formal analogy between the electric and mechanical problems, it is 
reasonable to assume that displacement fluctuations of a {\it single}
 bead do not provide
a direct measure of the bulk rheological properties whereas correlated 
fluctuations 
of {\it two} beads do, provided the beads are far enough apart.  
Indeed, we will 
find this to be the case, however, the vectorial nature of the elastic problem 
leads to some complications.

We begin by considering a single sphere of radius $a$ that perturbs the elastic medium
in which it is embedded out to a radius $a'$.  For $r > a'$, the medium is characterized
by bulk Lam\'{e} coefficients $\mu(\omega )$ and $\lambda(\omega )$.  For $r < a'$ 
the Lam\'{e} coefficients are $\bar{\mu}(\omega )$ and $\bar{\lambda}(\omega )$.
Elastic displacements ${\bf u}_{\rm inner}$ in the inner ($r<a'$) and ${\bf u}_{\rm%
outer}$ in the outer ($r>a'$) regions satisfy the equation
\begin{equation} 
\label{force-balance}
\mu \nabla^2 {\bf u} + \lambda {\bf \nabla} \left( {\bf \nabla} \cdot {\bf u} 
\right) = 0,
\end{equation}
where the Lam\'{e} constants take their appropriate values in each
region.  Putting the force applied to the sphere in the $\hat{z}$ 
direction,
the most general solution for ${\bf u}$ is
\begin{eqnarray} 
\label{solution}
{\bf u} &=& \frac{ a A}{r} \left(  \nu \hat{r} \cos \theta + \hat{z} \right) 
 + \frac{ a^3 B}{r^3} \left( 3 \hat{r} \cos \theta - \hat{z} \right) + \\ 
\nonumber
  &  &  + C \hat{z} + \frac{ D r^2}{a^2} \left( \sigma \hat{r} \cos \theta - 
\hat{z} \right) 
\end{eqnarray}
where $A,B,C$, and $D$ are constants.  The constants $\nu $ and $\sigma $ 
depend only on the local Lam\'{e} constants.  The 
solution for the strain field as written in Eq.~(\ref{solution}) includes  
a superposition
of a part that decays with distance as $1/r$ and a dipolar term.  These first two 
terms are accompanied by two other solutions: a constant shift and a 
term growing with distance from the sphere. The later two terms cannot 
occur in $\bu_{\rm outer}$
as that field must go to zero at large distances from the sphere. Thus in
the bulk solution for the strain we have two undetermined constants ($A$ and $B$)
while the inner solution ($\bu_{\rm inner}$)
has four undetermined constants.  The application of the boundary
conditions at infinity has reduced the problem to finding six constants.  
Due to the rigidity of the sphere, the displacement is
fixed at its surface. This boundary condition contributes two more constraints.
The remaining four conditions 
come from strain field continuity at the interface of 
the two elastic media ($r = a'$) and the continuity of the two components 
of the 
stress tensor at that interface, $\sigma_{rr}$ and $\sigma_{r \theta }$.  The
problem is now completely determined.

The complete solution shows that the self-component, $\alpha_{ij}^{11}$,
of the compliance tensor depends in a complex way on the local Lam\'{e}
coefficients $\bar{\mu}$ and $\bar{\lambda}$.  Thus as in the electrical case, 
fluctuations of a single bead will not yield reliable measurements of bulk 
rheology unless $a'/a - 1 \ll 1$ or $\bar{\mu}$, 
$\bar{\lambda}$ do not differ significantly from $\mu$,  $\lambda$.

To compute the cross component, $\alpha^{21}_{ij}(\omega )$, of the 
compliance tensor
relating displacements of bead 2 to forces on bead 1, we observe that bead 
2 will follow the displacement field produced by bead 1 at separations $r$ 
large compared to $a$.  Thus $\alpha^{21}_{ij}$ is simply the coefficient 
of $F_j^1$ in the displacement field of bead 1.  At large $r$, only the 
first term in Eq.~(\ref{solution}) survives.  
The coefficient $A$ in this term is determined by a global property of 
the stress field
\begin{equation}
\label{stress}
F_z = \oint ds_j  \sigma_{j z}, 
\end{equation}
where the integral is over any closed surface surrounding the sphere.  
Only the $1/r$ part of the displacement field contributes to this integral.  From 
this constraint we can calculate $A_{\rm outer}$, the coefficient of the the 
first term in Eq.~(\ref{solution}) in the outer region ($r>a'$).  This 
coefficient is linear in $F_z$.  From this, we find that the compliance tensor 
can be decomposed into parts, $\alpha_\|$, parallel to the vector ${\bf r}$ 
separating the two beads and, $\alpha_\perp$, perpendicular to 
${\bf r}$:  $\alpha_{ij}^{21}(\omega ) = \alpha_\| \hat{r}_i \hat{r}_j + %
\alpha_\perp \left( \delta_{ij} -  \hat{r}_i \hat{r}_j \right) $ with
\begin{eqnarray}
\label{ans-parallel}
\alpha_{\|} &=& \frac{ 1}{4 \pi r \mu(\omega )} \\
\label{ans-perp}
\alpha_{\perp} &=& \frac{ 1}{8 \pi r \mu(\omega )} \frac{ \lambda(\omega)  + 
3 \mu(\omega )}{\lambda(\omega) + 2 \mu(\omega )}.
\end{eqnarray}
Thus fluctuations parallel to the separation vector depend only on 
the shear modulus, $\mu(\omega ) = G(\omega )$, whereas those perpendicular
to the line of centers depend on both $\lambda $ and $\mu$.  In the incompressible
limit, $\alpha_\perp/ \alpha_\| = 1/2$, which is identical to the ratio of the 
parallel and perpendicular diffusivities of two spheres with (incompressible)
hydrodynamic interactions\cite{Batchelor:76}, in agreement with 
recent experimental
results on two-point microrheology in a viscous liquid\cite{Crocker:00}.
The experimental determination of this 
ratio in viscoelastic materials can be used to test for compressibility effects 
at the frequencies relevant to the experiment. 

The combination of single-particle and 
two-particle position correlations provide data about both the local 
environment of the probe particle and the bulk material. To test these ideas
we suggest that two particle position correlations should be measured at smaller
particle separations where $\alpha \nsim 1/r$. Correlations should then be 
sensitive to the particle's local environment.

We would like to thank J.C.~Crocker for communicating unpublished results and 
for many useful discussions including insight into the two-point correlation 
technique.  We would also like to thank R.D.~Kamien, F.C.~MacKintosh, and
A.G.~Yodh for helpful discussions. This work was supported in part by the
NSF MRSEC Program under grant No. DMR96--32598.

\end{document}